\RequirePackage{ifpdf}
\documentclass[hyper]{JHEP3}

\pdfoutput=1
\usepackage{epsfig}
\usepackage{cite}
\usepackage{graphicx}
\usepackage{subfigure}
\usepackage{inputenc}
\usepackage{amsmath}
\usepackage{comment}

\newcommand{\beq}{\begin{equation}}
\newcommand{\eeq}{\end{equation}}
\newcommand{\beqn}{\begin{eqnarray}}
\newcommand{\eeqn}{\end{eqnarray}}
\title{Limits on Electroweak Instanton-Induced Processes \\ 
with Multiple Boson Production
}

\author{Andreas Ringwald$^{a}$, Kazuki Sakurai$^{b}$,  Bryan R.~Webber$^{c}$\\
$^a$Deutsches Elektronen-Synchrotron DESY, 
Notkestra{\ss}e 85, D-22607 Hamburg, Germany\\
$^b$Institute of Theoretical Physics, Faculty of Physics, University of Warsaw, 
\\~ul.~Pasteura 5, PL--02--093 Warsaw, Poland\\
$^c$University of Cambridge, Cavendish Laboratory, J.J.\ Thomson Avenue, Cambridge, UK\\
E-mail: \email {andreas.ringwald@desy.de}, \email{kazuki.sakurai@fuw.edu.pl}, \email{webber@hep.phy.cam.ac.uk}
}

\preprint{Cavendish-HEP-18/15 \\ DESY 18-169}

\abstract{
Recently, the CMS collaboration has reported their search for electroweak instanton-like processes 
with anomalous $B+L$ violation assuming multi-fermion but zero-boson final states.
On the other hand, many theoretical studies suggest that anomalous $B+L$ processes may have an observably large
production rate only if their final state contains a large number of electroweak gauge bosons.
In this paper, we review the state-of-the-art of the predictions of electroweak instanton-induced processes
and compare collider signatures of 
zero- and multi-boson events of anomalous $B+L$ violation at the LHC.
An upper limit on the cross-section for the multi-boson process is derived 
by recasting the CMS analysis. 
}

\begin{document} 

\section{Introduction}
\label{sec:intro}

The discovery of the Higgs boson at the LHC has marked the completion of the Standard Model (SM) of electroweak and strong interactions. Moreover, so far, 
no compelling evidence for physics beyond the SM has been observed at the LHC. In fact, the perturbative SM predictions for 
hard, short distance dominated scattering processes are all verified experimentally to a remarkable precision. 

Theory predicts, however, that there are also hard scattering processes in the SM whose amplitudes are fundamentally 
non-perturbative. Their existence follows from axial anomalies~\cite{Adler:1969gk,Bell:1969ts,Bardeen:1969md} 
in the SM, resulting in an anomalous violation of baryon plus lepton number, 
$B+L$, in electroweak interactions~\cite{tHooft:1976rip,tHooft:1976snw}.
They are induced by topological fluctuations of the non-Abelian SU(2)$_L$ gauge fields,  
notably electroweak instantons~\cite{Belavin:1975fg,Affleck:1980mp}.  
Anomalous $B+L$ violating processes are believed to be very important in the high temperature 
primordial plasma after the big bang~\cite{Kuzmin:1985mm,Arnold:1987mh,Ringwald:1987ej}
and to have therefore a crucial impact on the evolution of the baryon and lepton
asymmetries of the universe (see Ref.~\cite{Rubakov:1996vz} for a review).

A very interesting, albeit unsolved question is whether manifestations of such topological
fluctuations might be directly observable in high-energy scattering at present or 
future colliders. 
This question has been raised originally in the late eighties in~\cite{Ringwald:1989ee,Espinosa:1989qn}, 
but, despite a lot of effort, the actual
size of the cross-sections in the relevant, tens of TeV energy regime could not be determined.

Nevertheless, it has been established  (for reviews, see Refs.~\cite{Rubakov:1996vz,Mattis:1991bj,Tinyakov:1992dr})
that the inclusive cross-section for electroweak $(B+L)$-violating processes, 
\begin{equation}
q+q \, \to \, 7 \bar q + 3 \bar \ell + \sum_{n_{W,Z}} n_{W,Z} W(Z) + \sum_{n_h} n_h h ,
\label{eq:process}
\end{equation} 
can be written in  the form 
\begin{equation}
\label{eq:stot}
{\hat \sigma}_{\rm incl}^{\rm BV}
\equiv \sum_{n_W,n_h} {\hat \sigma}^{\rm BV}_{n_W,n_h} 
= \frac{P(\epsilon   )}{m_W^2} 
 \exp\left[ -
\frac{4\pi}{ \alpha_W } F
\left( \epsilon \right)
\right] \,,
\end{equation}
where
\begin{equation}
\epsilon \equiv \frac{\sqrt{\hat s}}{M_0}
\end{equation}
is the quark-quark centre-of-mass energy $\sqrt{\hat s}$ in terms of the scale 
\begin{equation}
\label{eq:m0}
M_0\equiv \sqrt{6}\pi\frac{m_W}{\alpha_W} \simeq 18.3\ {\rm TeV}\,,
\end{equation}
which is parametrically similar to the energy of the sphaleron, $E_{\rm sph}=f(m_h^2/m_W^2) \pi m_W/\alpha_W \sim 9$ TeV -- the static energy of a classically unstable saddle-point solution 
of the static bosonic field equations of the electroweak theory,
corresponding to the minimal barrier height between inequivalent vacua
with different values of the Chern-Simons topological index $N_{\rm CS}$~\cite{Manton:1983nd,Klinkhamer:1984di}.  
The `holy grail function' $F$ in the exponent is known in terms of an expansion in $\epsilon$, whose first
few terms are given by~\cite{Khlebnikov:1990ue,Khoze:1990bm,Mueller:1991fa,Diakonov:1993ur,Balitsky:1993xc}
\begin{eqnarray}
 \label{eq:holygrail}
F(\epsilon)&=&1-{9\over 8}\epsilon^{4/3}+
{9\over 16}\epsilon^2  
 - {3\over 32}\left(4-3{m_h^2\over m_W^2}\right)\epsilon^{8/3}\ln {1\over \epsilon}
+
{\mathcal O}(\epsilon^{8/3}\cdot {\rm const}) \,,
\end{eqnarray}
while the pre-exponential factor $P$ reads \cite{Khoze:1990bm,Ringwald:2002sw}\footnote{
  The difference in the energy dependence (i.e.~the powers of $\epsilon$)
  between Eq.~\eqref{eq:prefactor} and the formula in Ref.~\cite{Khoze:1990bm} 
  is due to the different number of fermions assumed.
  For the general case involving $n_f$ fermions, $P(\epsilon) \propto \epsilon^{\frac{14 + 5 n_f}{9}}$.
} 
\begin{eqnarray}
\label{cross-qfd}
P(\epsilon  )  
=
\frac{\pi^{15/2}}{1024} 
\ \left( \frac{3}{2}   \right)^{\frac{2}{3}} \,d^2
\
\left( \frac{4\pi}{\alpha_W}\right)^{7/2}
\,
\epsilon^{\frac{74}{9}}
\left[ 1 + \mathcal{O}\left( \epsilon^{2/3}\right)\right]
\,,
\label{eq:prefactor}
\end{eqnarray}
where $d\simeq 0.15$.

Obviously, in the energy region $m_W\ll \sqrt{\hat s}\ll M_0$, 
the inclusive cross-section for ($B+L$)-violation is 
unobservably small, due to the tremendous exponential suppression 
factor, $\exp (-4\pi/\alpha_W) \sim 10^{-162}$, typical for a quantum 
tunneling process. Nevertheless, it is exponentially growing in this energy region.  
Intriguingly, the characteristic scale for the exponential growth is $M_0$ and thus 
of the order of the sphaleron energy. One may interpret this as a hint that the process proceeds 
via an intermediate virtual sphaleron-like field configuration. This interpretation is also 
backed-up by the fact that the inclusive cross-section~(\ref{eq:stot})
is dominated by a semi-classically large number of $W(Z)$ and Higgs bosons 
\cite{Zakharov:1990dj}, 
\begin{equation}
\label{eq:mul}
\langle n_W \rangle =  {4\pi\over 
\alpha_W} \left( {3\over 8} \epsilon^{4/3} + {\cal{O}}\left(
\epsilon^{2}\right) \right) ,
\end{equation}
and 
\begin{equation}
\label{eq:mulH}
\langle n_h \rangle \simeq  {4\pi\over\alpha_W}{3\over 32} \epsilon^2 \;  ,
\end{equation}
in line with the expectation from the decay of a sphaleron-like intermediate state \cite{Aoyama:1986ej}. 
Unfortunately, however, 
the perturbative expansion of the holy-grail function $F$ in powers
of $\epsilon$ breaks down around $\epsilon \sim 1$ and therefore it is not 
possible to infer the size of the cross-section at centre-of-mass
energies relevant for the LHC. 

The multi-boson events such as Eq.~\eqref{eq:process}
lead to spectacular collider signatures.
According to Eq.~\eqref{eq:mul}, one expects ${\cal O}(30)$ electroweak bosons in an event on average.\footnote{
The number of Higgs bosons is only a few according to Eq.~\eqref{eq:mulH}.}
To study collider signatures of Eq.~\eqref{eq:process}, the
event generator {\tt HERBVI} \cite{HERBVI} was developed,
and the first thorough phenomenological study was carried out in Ref.~\cite{Gibbs:1994cw}.

After the end of the SSC project in the early nineties, the motivation for theorists to decide this question immediately 
dropped considerably and the research in this direction essentially
stopped.\footnote{Instead, then the
  theory~\cite{Moch:1996bs,Ringwald:1998ek},
  phenomenology~\cite{Ringwald:1999jb} and experimental
  searches~\cite{Adloff:2002ph,Chekanov:2003ww,H1:2016jnv} for the
  analogous instanton-induced processes in QCD were developed.}
However, a recent paper by Tye and Wong \cite{Tye:2015tva}, presenting
qualitative arguments in favour of observable rates for electroweak
($B+L$)-violation, revived the interest in it and motivated
phenomenological~\cite{Ellis:2016ast,Brooijmans:2016lfv,Ellis:2016dgb,
Spannowsky:2016ile,Funakubo:2016xgd,Cerdeno:2018dqk,Jho:2018dvt}
and experimental studies.
In particular, the CMS collaboration 
has recently reported their results 
of searching for signatures of electroweak instanton-induced processes at the LHC \cite{cms}.   

There is, however, a remarkable difference between the processes 
considered in the CMS analysis and the virtual sphaleron-like processes
discussed in the literature and briefly summarised above.
Following earlier studies \cite{Ellis:2016ast}
and for simplicity, CMS postulated the zero-boson process
\beq
q q \,\to\, n_q q \,+\, 3 \ell ~~~ 
\label{eq:zero-boson-proc}
\eeq
where $n_q = 7$, 9 or 11,
and $q/\ell$ represent quark/lepton or anti-quark/anti-lepton
depending on how the fermion lines are contracted with an effective vertex
in the instanton background with topological charge $\Delta N_{\rm CS}= \pm 1$.
In the CMS analysis, the event generator {\tt BaryoGEN} \cite{BaryoGEN} was used 
to simulate the process in Eq.~\eqref{eq:zero-boson-proc}.

In this paper we investigate the difference in collider signatures between
multi- and zero-boson process (i.e.~Eq.~\eqref{eq:process} and Eq.~\eqref{eq:zero-boson-proc})
using the {\tt HERBVI} and {\tt BaryoGEN} event generators, respectively.
After comparing various distributions in section \ref{sec:highmulti}, 
we recast the original CMS analysis and derive the cross-section upper limit on the
sphaleron-like multi-boson process in section \ref{sec:recast}.
Section \ref{sec:conclusions} is devoted to conclusions.

\medskip
\section{Event simulation}
\label{sec:evsim}

It has been suggested that the steeply-rising event rate of
sphaleron-like processes can be modelled by a sharp threshold behaviour 
as a function of the partonic centre-of-mass energy, $\sqrt{\hat s}$ \cite{Ringwald:1990qz}.
We therefore postulate the following partonic cross-section:
\beq
\hat \sigma( \sqrt{\hat s} ) \,=\, \frac{p_{\rm sph}}{m_W^2} \Theta( \sqrt{\hat s} - E_{\rm sph} ),
\label{eq:xsec}
\eeq
where $\Theta(x)$ is the Heaviside theta function; $\Theta(x) = 1$ for $x \ge 0$ and 0 otherwise,
and $E_{\rm sph}$ is the partonic threshold energy. 
We take $E_{\rm sph} = 9$ TeV as the default value for our simulation
(later varying it by $\pm1$ TeV), motivated by the fact that 
the sphaleron barrier height with $m_h = 125$ GeV is given by $\sim
9.1$ TeV \cite{Manton:1983nd}.  
In Eq.~\eqref{eq:xsec}, the cross-section is normalised by $1/m_W^2$ up to the dimensionless parameter $p_{\rm sph}$, which controls the overall cross-section.
The same parametrisation is used in the CMS analysis \cite{cms}.  

The event samples for zero- and multi-boson final states are generated
by {\tt BaryoGEN} \cite{BaryoGEN} and {\tt HERBVI} \cite{HERBVI},
respectively, for $pp$ collisions at centre-of-mass energy $\sqrt{s} = 13$ TeV.
Since {\tt BaryoGEN} only generates partonic final states from the hard process,
the sample is passed to {\tt Pythia 8} \cite{pythia} to simulate the
parton shower and hadronization. {\tt HERBVI} is a hard process
generator interfaced to the Fortran {\tt HERWIG} generator~\cite{Corcella:2000bw}.
For this study we use the last Fortran version {\tt HERWIG 6.521}~\cite{Corcella:2002jc}. 

The above hadronic samples are then passed to {\tt Delphes}~\cite{deFavereau:2013fsa} to simulate the detector response.
We tune the parameters in {\tt Delphes} in such a way that it emulates the CMS analysis \cite{cms} as accurately as possible.
Jets are reconstructed using the anti-$k_T$ algorithm with cone size 0.4.
Leptons (electrons and muons) and photons are required to be isolated from neighbouring energy activity 
within $\Delta R \equiv \sqrt{\Delta \phi^2 + \Delta \eta^2} < 0.4$ (muons) or 0.3 (electrons and photons).
We require the scalar sum of the transverse momenta $p_T$ of the particles within this cone (excluding the targeted object itself (lepton or photon)) 
to be less than 15\% (muons), 10\% (electrons) or 12\% (photons) of the $p_T$ of the targeted object.
We demand $p_T > 30$ GeV for all objects (jets, leptons and photons)
and $|\eta| < 5$ (jets), $< 2.4$ (muons) and $<2.5$ (electrons and photons).

\FIGURE[t]{
 \centering
  \includegraphics[width=0.65\textwidth,clip]{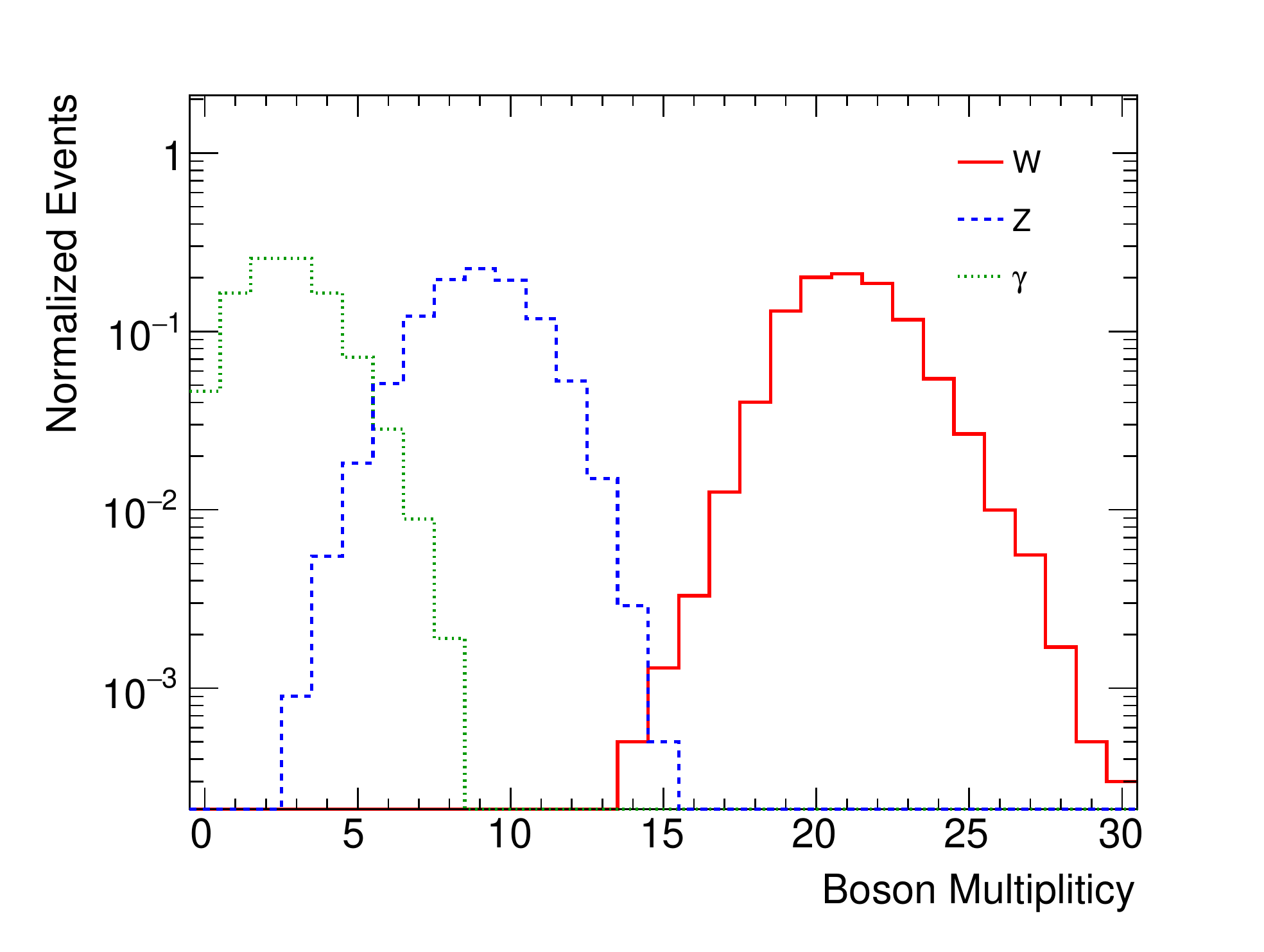}
     \vspace{-3mm}
  \caption{\label{fig:NB} 
  Normalised distributions of the number of $W$, $Z$ and $\gamma$ generated by {\tt HERBVI}.
  }
}

\section{Zero vs high boson multiplicity events}
\label{sec:highmulti}

We begin this section by showing 
the boson multiplicity distributions 
obtained by {\tt HERBVI} in Fig.~\ref{fig:NB}.
The distributions are obtained as follows.
First, the gauge boson multiplicity distribution has been
calculated in the symmetric phase using the leading-order matrix element formula 
given in \cite{Ringwald:1989ee, Espinosa:1989qn, Gibbs:1994cw}.
The distributions of $Z$ and $\gamma$ are obtained by transforming the neutral $W$ bosons into $Z$ and $\gamma$
with probabilities $\cos^2 \theta_W$ and $\sin^2 \theta_W$, respectively.
One can see in the figure that the $W$ boson multiplicity peaks around 20.
One the other hand, the central value of the $Z$ multiplicity is about
a factor of 2 smaller.
The number of photons per event is rather modest but 2 or 3 photons are expected on average.
We do not show the Higgs boson multiplicity distribution because it is negligible compared to that of
gauge bosons.

We now compare various distributions obtained for the events with zero- ({\tt BaryoGEN}) and multi- ({\tt HERBVI}) boson final states in Fig.~\ref{fig:multi}.
In this and the rest of the figures, the blue-dashed and red-solid histograms correspond to the distributions for the zero- and multi-boson final state events, respectively.

\FIGURE[h]{
 \centering
    \includegraphics[width=0.5\textwidth,clip]{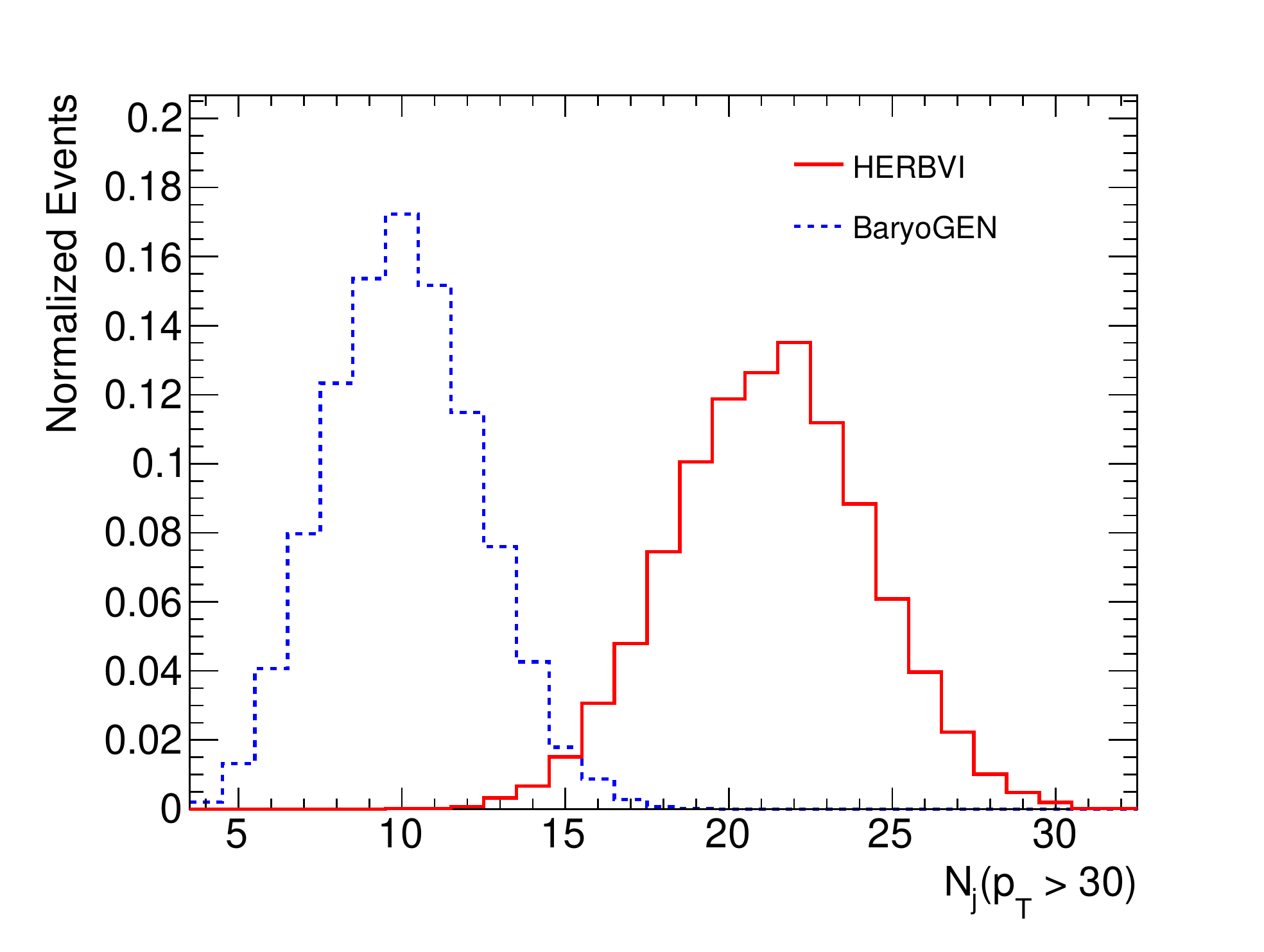}
    \hspace{-3mm}
    \vspace{-3mm}
    \includegraphics[width=0.5\textwidth,clip]{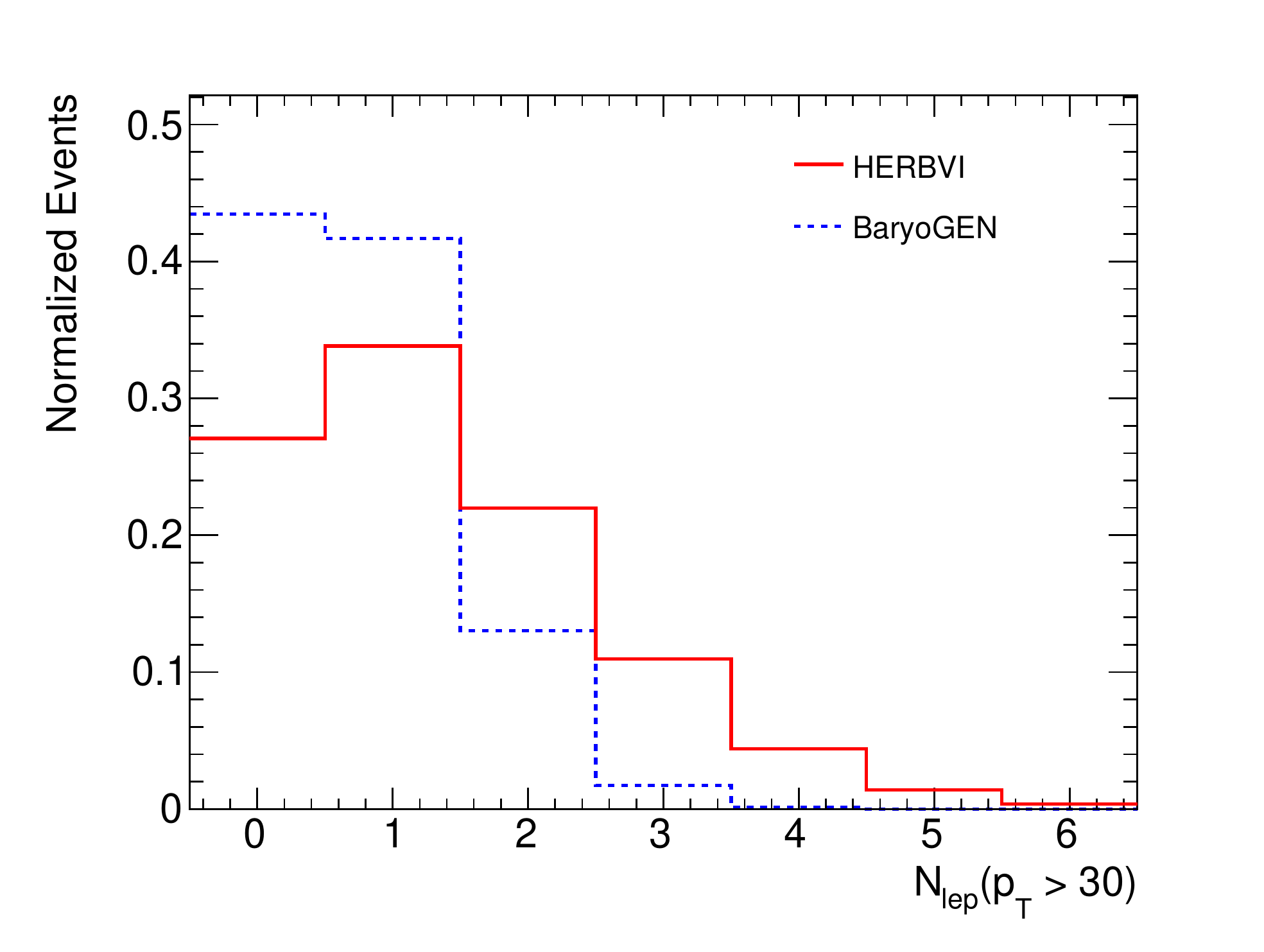}
    \hspace{-3mm}
    \includegraphics[width=0.5\textwidth,clip]{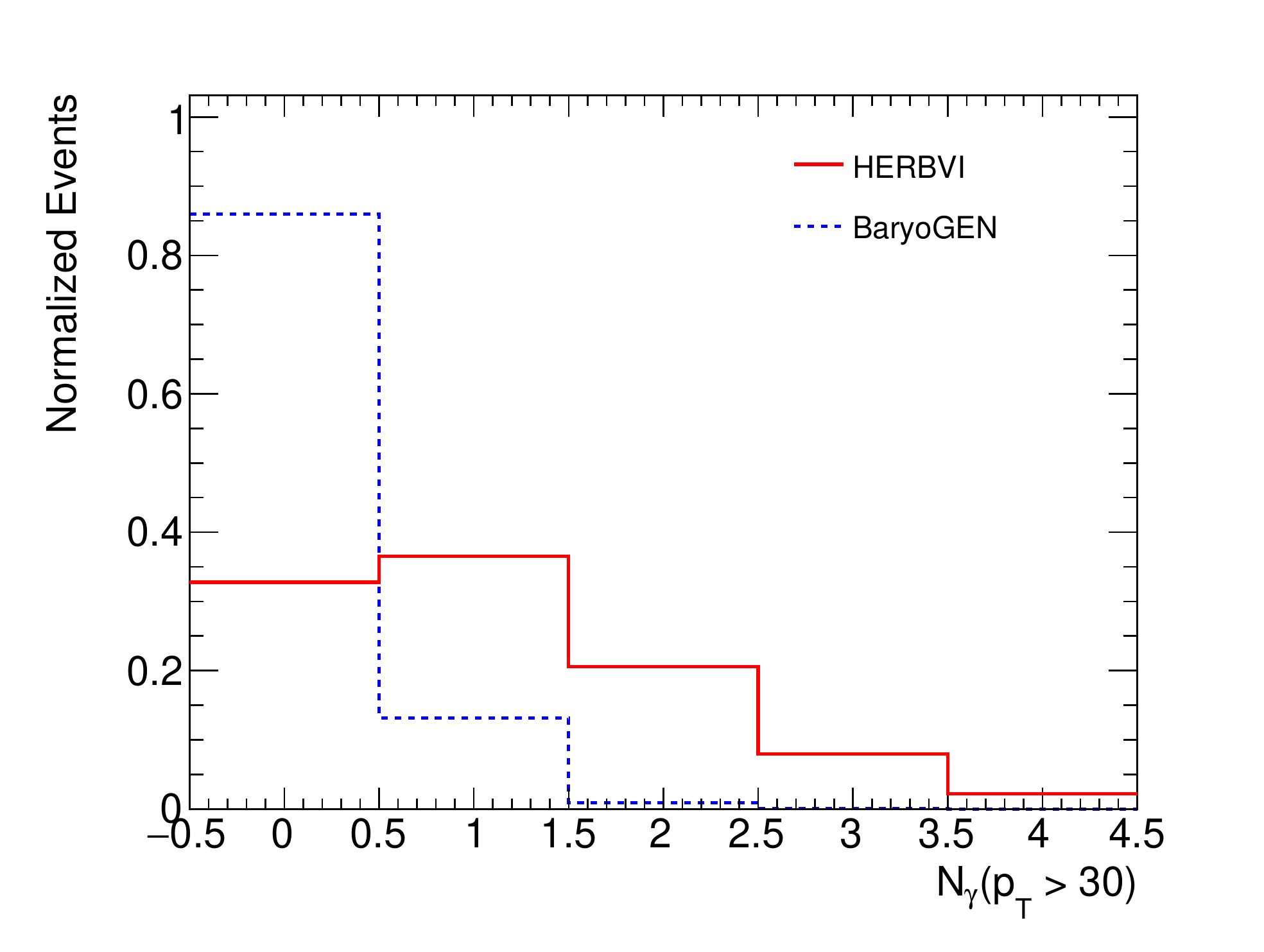}
     \vspace{-3mm}
     \caption{\label{fig:multi}
    The multiplicities of final-state objects with $p_T > 30$ GeV; jets (top), leptons (bottom left),
    photons (bottom right).
    The red-solid and blue-dashed histograms correspond to 
    the multi- and zero-boson events generated by {\tt HERBVI} and {\tt BaryoGEN}, respectively.
}
}

The top panel shows the jet multiplicity distributions.
As expected, the distribution for zero-boson final states peaks around
$N_j = 10$, since 7, 9 or 11 quarks are present in the final states
coming from the hard interaction, although some extra jets can arise
from initial- and final-state QCD radiation.
On the other hand, $N_j$ peaks around 22 in the multi-boson case.
The extra jets are coming mainly from the decay of heavy electroweak
bosons.
It is interesting that $~99$\% of the multi-boson events have $N_j \ge 15$.
In such a high jet multiplicity region the Standard Model background is extremely low. 

The lepton multiplicities are compared in the bottom left panel of Fig.~\ref{fig:multi}.
As can be seen, the number of leptons in an event is much larger for the multi-boson process,
since the extra leptons are produced in the decay of the heavy bosons.
While only $\sim 20$\% of the zero-boson events have $N_{\ell} > 1$,
this rate is $\sim 40$\% for the multi-boson events.

A large difference is also found in the photon multiplicity distributions, shown in the bottom right panel of Fig.~\ref{fig:multi}.
For the zero-boson events, photons are produced only from QED radiation or decays of hadrons,
while for the multi-boson process, high-$p_T$ photons may be produced also from the primary hard interaction.
For the zero-boson case, $\sim 85 \%$ of events do not have any such photons and
the probability of having more than one photon is extremely small ($< 1 \%$). 
On the other hand, the photon multiplicity distribution peaks around $N_\gamma = 1$
for the multi-boson case and has a tail towards the higher multiplicity region, $N_\gamma = 2, 3$.
For example, the rate of having three high-$p_T$ photons is as large as $\sim 10\%$.

\medskip
\section{Limits on sphaleron process}
\label{sec:recast}

The CMS collaboration has recently carried out an analysis searching for mini black holes 
and sphalerons that could be produced by the 13 TeV proton-proton collisions at the LHC \cite{cms}. 
The dataset used was collected in 2016 and corresponds to an integrated luminosity of 35.9 fb$^{-1}$.  
The events are recorded if they pass the trigger requirement of $H_T > 800$ or 900 GeV,
depending on the period when the data was collected,
where $H_T$ is defined as the scalar sum of the $p_T$ of all jets reconstructed at the stage of the High Level Trigger.

The CMS analysis adopts a simple cut-and-count method. 
The cuts are imposed on the two variables,  
$N(p_T > 70)$ and $S_T$,
where $N(p_T > 70)$ is the number of all reconstructed objects (jets, leptons and photons) with $p_T > 70$ GeV
and $S_T$ is the scalar sum of the transverse momenta of all reconstructed objects with $p_T > 70$ GeV including 
the missing transverse energy;
$S_T \equiv \sum_i^{p_T > 70 \rm GeV} p^{(i)}_T$, ($i = $ jets, leptons, photons and $E_T^{\rm miss}$).

The signal regions are labelled by ($N_{\rm min}$, $S_T^{\rm min}$),
where the events are counted if 
the conditions $N(p_T > 70) \ge N_{\rm min}$ and $S_T > S_T^{\rm min}$
are both satisfied.
The following ranges were used for signal regions; $N_{\rm min} = 3, ..., 11$ and $3.8 < S_T^{\rm min}/{\rm TeV} < 8$.

CMS performed a careful estimation of the Standard Model background.
The dominant background comes from QCD multi-jets, 
followed by $V + {\rm jets}$, $\gamma + {\rm jets}$ and $t \bar t$ production.
By comparing the number of observed events with the expected
background contribution in each signal region,
CMS estimated a model-independent 95\% CL upper limit on the extra contribution to the signal region.
We refer to this bound as $N_{\rm obs}^{{\rm max}(a)}$, where $a$ labels the signal region in question.
CMS also estimated the ``expected'' upper limit, $N_{\rm exp}^{{\rm max}(a)}$,
by assuming that exactly the same (fractional) number of events predicted by the background
was observed.

Our procedure for setting limits is as follows.
First, we estimate the contribution from sphaleron events to the signal region $a$, $N_{\rm sph}^{(a)}$,
using the {\tt BaryoGEN} or {\tt HERBVI} Monte Carlo simulation.
These signal contributions are then compared with the corresponding expected upper limits to 
identify the ``most sensitive'' signal region, $a^*$, that gives the largest value for $N_{\rm sph}^{(a)}/N_{\rm exp}^{{\rm max}(a)}$, i.e.~$\forall a; N_{\rm sph}^{(a^*)}/N_{\rm exp}^{{\rm max}(a^*)} \ge N_{\rm sph}^{(a)}/N_{\rm exp}^{{\rm max}(a)}$.
We then compare the signal contribution with the observed upper limit 
only at the most sensitive signal region.
We exclude the signal hypothesis if $N_{\rm sph}^{(a^*)}/N_{\rm obs}^{{\rm max}(a^*)} > 1$.

\begin{table*}[t!]
\begin{center}
\def\arraystretch{1.4}
\begin{tabular}{|c|c||c|c|c|c|c|}
\hline
\multicolumn{2}{|c||}{ $E_{\rm sph}$ [TeV] } & 8 & 8.5 & 9 & 9.5 & 10 \\ \hline \hline
            & $(N_{\rm min}, S_T^{\rm min}\,[\rm TeV])^*$ & (11, 4.2) & (11, 4.2) & (11, 4.2) & (11, 4.2) & (11, 4.2) \\ \cline{2-7} 
multi-boson & $\epsilon^{(a^*)}$ [\%] & 94.8 & 97.5 & 99.2 & 99.6 & 99.9 \\ \cline{2-7}
            & $N_{\rm obs}^{{\rm max} (a^*)}$ & 3.0 & 3.0 & 3.0 & 3.0 & 3.0 \\ \hline 
\hline
            & $(N_{\rm min}, S_T^{\rm min}\,[\rm TeV])^*$ & (9, 5.4) & (9, 5.6) & (9, 5.6) & (8, 6.2) & (8, 6.2) \\ \cline{2-7} 
Zero Boson  & $\epsilon^{(a^*)}$ [\%] & 37.7 & 40.5 & 45.3 & 50.5 & 57.5 \\ \cline{2-7}
            & $N_{\rm obs}^{{\rm max} (a^*)}$ & 6.9 & 5.8 & 5.8 & 3.0 & 3.0 \\ \hline 
\end{tabular}
\caption{\label{tb}
The most sensitive signal region ($a^* = (N_{\rm min}, S_T^{\rm min})^*$),
the signal efficiency ($\epsilon^{(a^*)}$) at signal region $a^*$,
and the observed upper limit for signal region $a^*$ ($N_{\rm obs}^{\rm max {(a^*)}}$)
are shown for different values of the partonic threshold energy $E_{\rm sph}$.
}
\end{center}
\end{table*}

In Table \ref{tb}, we show the most sensitive signal region ($a^* = (N_{\rm min}, S_T^{\rm min}\,[\rm TeV])^*$),
the signal efficiency ($\epsilon^{(a^*)}$) at signal region $a^*$,
and the observed upper limit for signal region $a^*$ ($N_{\rm obs}^{\rm max {(a^*)}}$)
for different values of the partonic threshold energy $E_{\rm sph}$.
The signal efficiency is defined as the fraction of events that satisfy the condition employed by the signal region.
In particular, it satisfies the relation
$N_{\rm sph}^{(a)} = \sigma \cdot \epsilon^{(a^*)} \cdot L_{\rm int}$,
where 
$L_{\rm int}$ is the integrated luminosity
and
$\sigma$ 
is the inclusive hadronic cross-section 
obtained by convoluting the partonic cross-section in Eq.~\eqref{eq:xsec} with the parton distribution functions (PDFs).
 In Fig.~\ref{fig:xsec} we show the inclusive hadronic cross-section
 with $p_{\rm sph} = 1$ in Eq.~\eqref{eq:xsec},
where the {\tt CT10} LO PDF set \cite{ct10} was used following the CMS analysis \cite{cms}. 
%
\FIGURE[h]{
 \centering
    \includegraphics[width=0.6\textwidth]{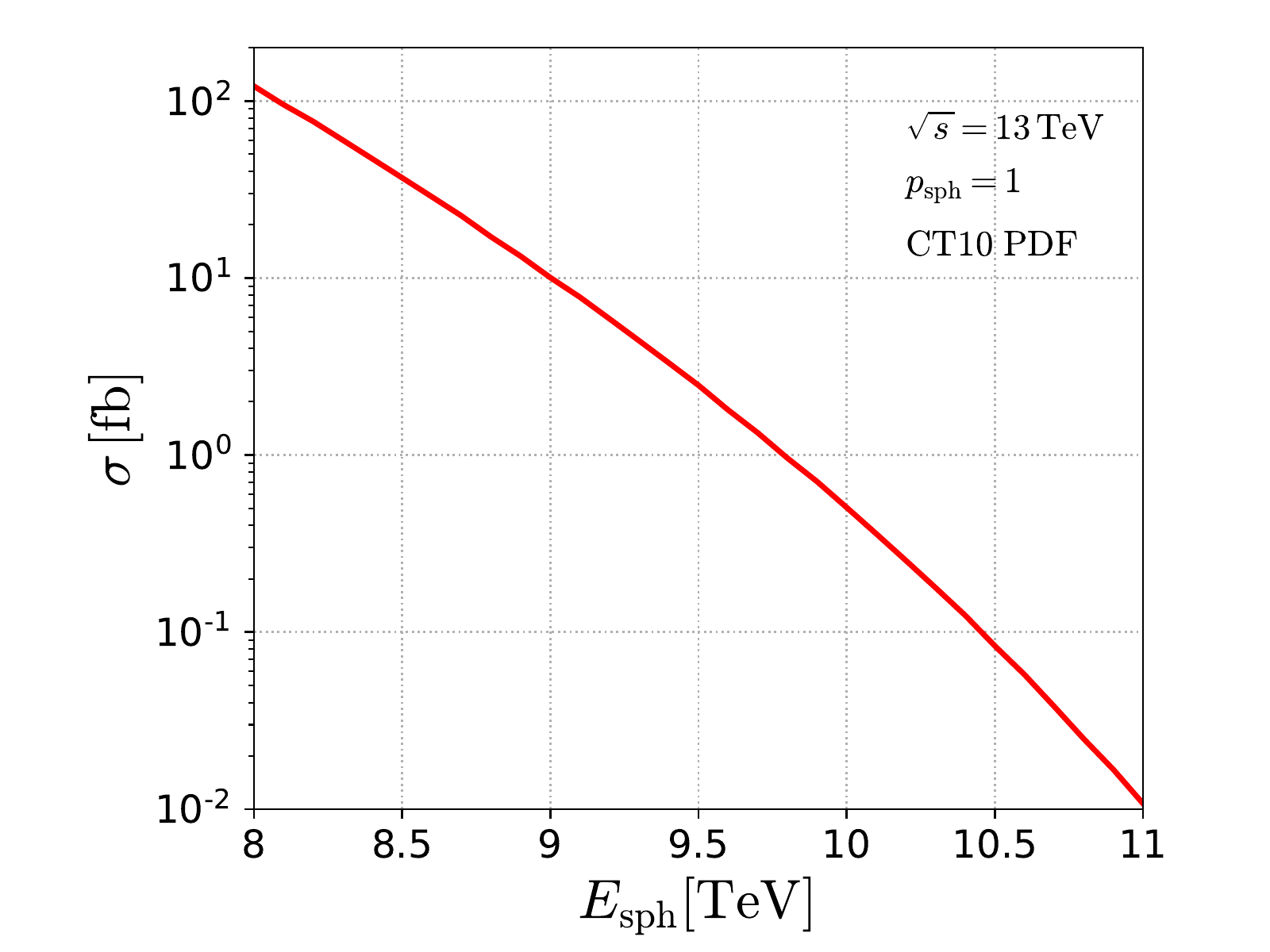}
    \caption{\label{fig:xsec}
    The 13 TeV inclusive hadronic cross-section for sphaleron production with $p_{\rm sph} = 1$
    as a function of $E_{\rm sph}$ in Eq.~\eqref{eq:xsec}, obtained
    with the {\tt CT10} LO PDF set.
}
}

\FIGURE[h]{
 \centering
    \includegraphics[width=0.5\textwidth,clip]{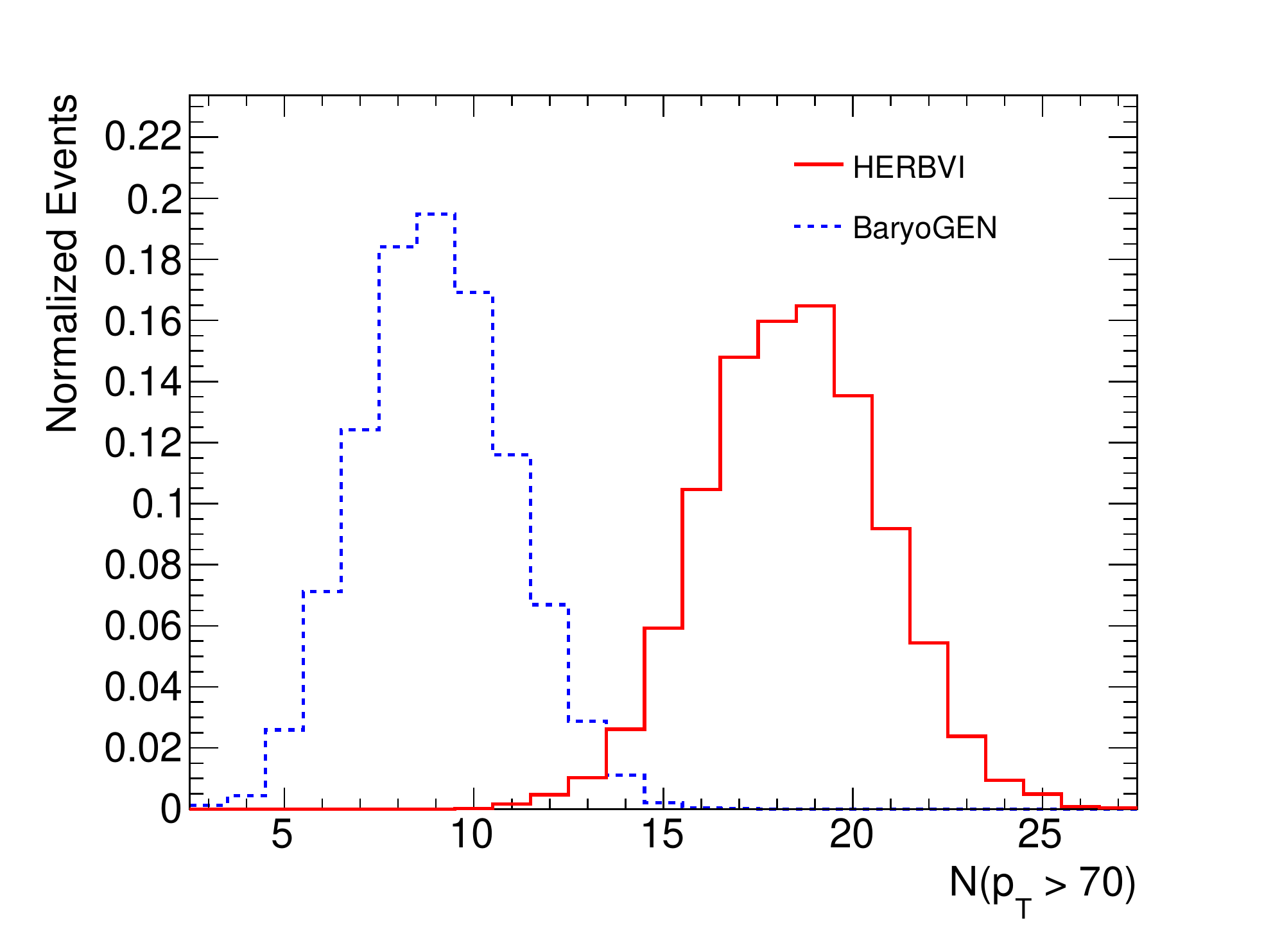}
    \hspace{-5mm}
    \includegraphics[width=0.5\textwidth,clip]{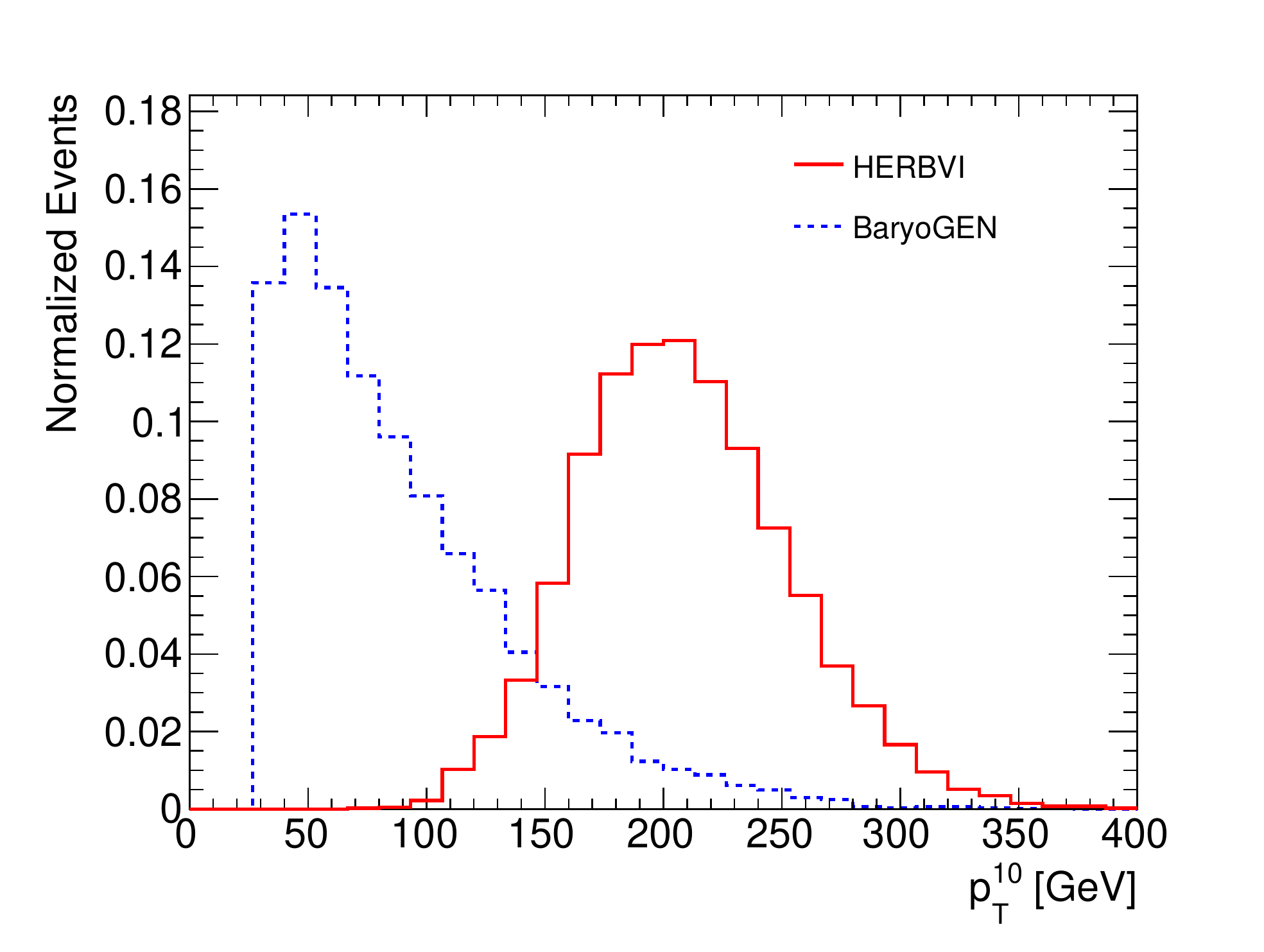}
    \vspace{-5mm}    
    \caption{\label{fig:pt10}
Left panel: the inclusive multiplicity containing jets, leptons and photons with $p_T > 70$ GeV.  
Right panel: the transverse momentum distributions of the 10th hardest jet. 
The red-solid and blue-dashed histograms correspond to {\tt HERBVI} and {\tt BaryoGEN} event generators.
    }
}

In Table \ref{tb}, we see that $N_{\rm min} = 11$, $S_T^{\rm min} = 4.2$ TeV is selected 
for the most sensitive signal region throughout the range of $E_{\rm sph}$ for the multi-boson case,
whilst $N_{\rm min} = 8, 9$ and $S_T^{\rm min} \in [5.4, 6.2]$
are chosen for the most sensitive signal regions depending on $E_{\rm sph}$ for the zero-boson case. 
The tendency is clear that the multi-boson process favours the high multiplicity and small $S_T$ regions,
while the zero-boson process prefers the low multiplicity and large $S_T$ regions in comparison.

To understand this, we first look at the $N(p_T > 70)$ distributions shown in the left panel of Fig.~\ref{fig:pt10}.
One can see that $N(p_T > 70)$ for the zero-boson case peaks around 8,
which is significantly lower than the typical multiplicity of the primary partons, $N_0 \sim 12$, produced from the hard interaction.
In fact, only half of the events have $N(p_T > 70) \ge 8$.
The reason for this may be understood as follows.
Since the primary partons from the zero-boson process are only quarks and leptons,
$N_0 \sim 12$ should be interpreted as the {\it maximum} of the $N(p_T > 70)$ distribution rather than the average.
In fact, among $N_0$ partons at most three of them may be neutrinos, which will of course not be counted in $N(p_T > 70)$.
Furthermore, the $p_T$ threshold of 70 GeV may have a large impact on the lower $p_T$ objects.
To see this we show the $p_T$ distributions of the 10th hardest object in the event 
in the right panel of Fig.~\ref{fig:pt10}. 
We see from this plot that the distribution for the zero-boson case 
peaks near the lowest $p_T$ bin, and only $\sim 40\%$ of the 10th hardest objects have $p_T > 70$ GeV.
This is not surprising because large differences in $p_T$ are expected between the hardest and softest 
primary objects.
For the multi-boson case, this effect is much smaller because 
the event typically contains many more than 10 primary objects, 
and the difference in $p_T$ between the hardest and 10th hardest objects is milder. 

As seen in the left panel of Fig.~\ref{fig:pt10}, 
the lower tail of $N(p_T > 70)$ barely reaches 11 for the multi-boson case.
This means the the multi-boson signal need not pay any price 
for the condition $N(p_T > 70) \ge N_{\rm min}$, 
since CMS varies $N_{\rm min}$ only up to 11 in their signal regions.
In fact, the multi-boson signal efficiencies are always $\gtrsim 95 \%$ 
in the most sensitive signal region $(N_{\rm min}, S_T^{\rm min})^* = (11, 4.2\,{\rm TeV})$.
The larger the $N_{\rm min}$, the lower the Standard Model background,
and $N_{\rm min} = 11$ is therefore selected 
throughout $E_{\rm sph}$
in the most sensitive signal region for the multi-boson case.

\begin{figure}[t!!]
\begin{center}
    \includegraphics[width=0.6\textwidth]{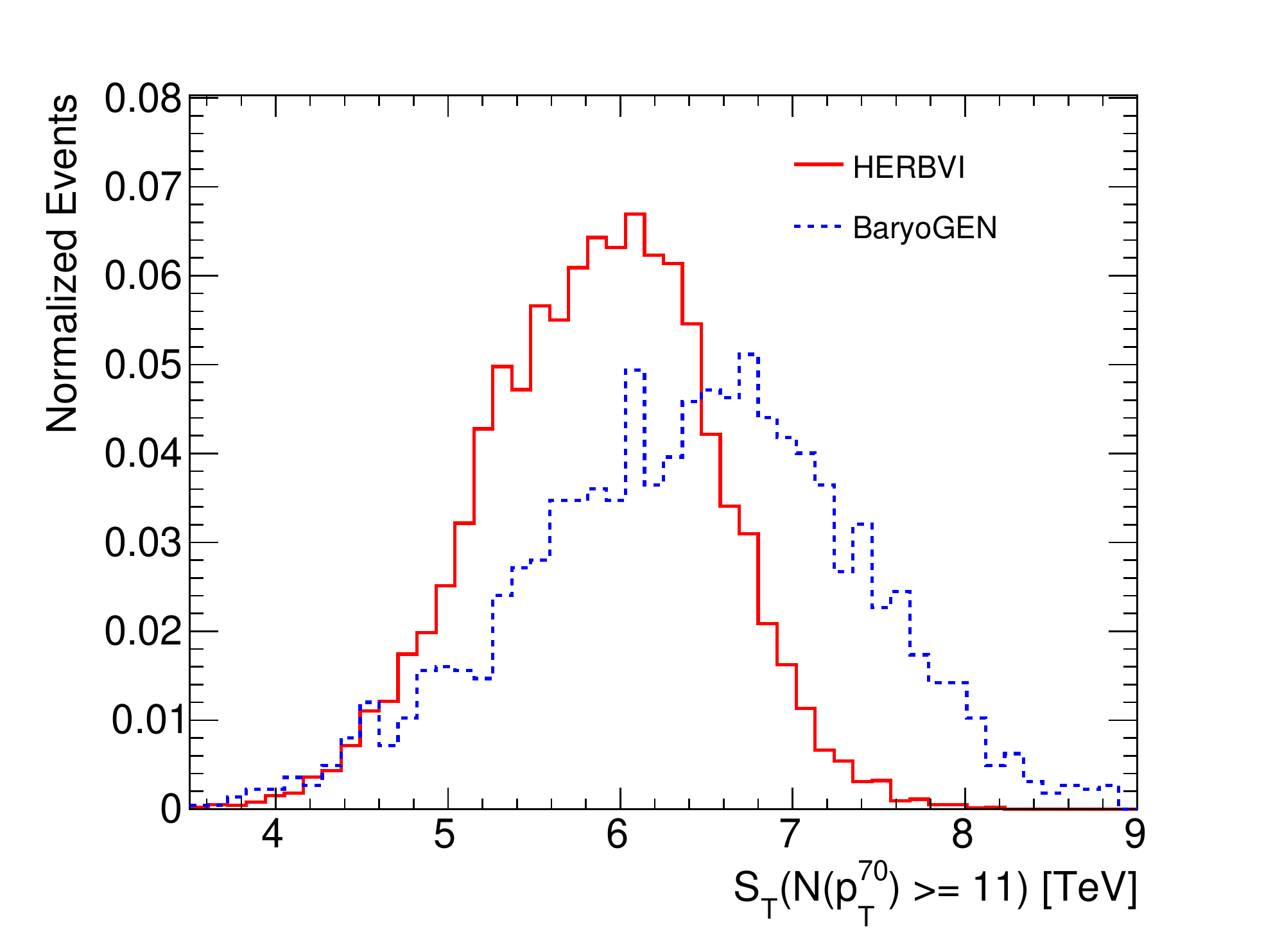}
    \caption{ 
    The $S_T$ distribution with the condition $N(p_T > 70) \ge 11$.
    The red-solid and blue-dashed histograms correspond to {\tt HERBVI} and {\tt BaryoGEN} event generators.    
    }
    \label{fig:ST}
\end{center}
\end{figure}

We now look at the $S_T$ distributions in Fig.~\ref{fig:ST} with $N_{\rm min} = 11$. 
We see that the values of $S_T$ are generally quite high.
The distribution peaks around 6 TeV for the multi-boson process
and 7 TeV for the zero-boson one.
This is expected because the $S_T$ variable is
designed to reflect the total energy of the event, $E_{\rm sph} \sim 9$ TeV.
One can also see that $S_T$ is generally lower in the multi-boson case
compared to the zero boson case.
This is because more objects are rejected by the $p_T > 70$ GeV cut.
For the multi-boson process, 
the number of primary partons (including those produced from heavy boson decays)
may be estimated as $N_0 \sim {\cal O}(70)$
assuming $n_B \sim {\cal O}(30)$ and each boson decays into two partons.
Therefore, the total event energy is distributed among $N_0$ partons,
yielding a large chance for each one to be rejected by the $p_T > 70$ GeV cut.
However, the impact of this on $S_T$ is not very large.
More than $98 \%$ of the events are accepted by $S_T > 4$ TeV for both zero- and multi-boson samples.

\begin{figure}[t!!]
\begin{center}
    \includegraphics[width=0.52\textwidth,clip]{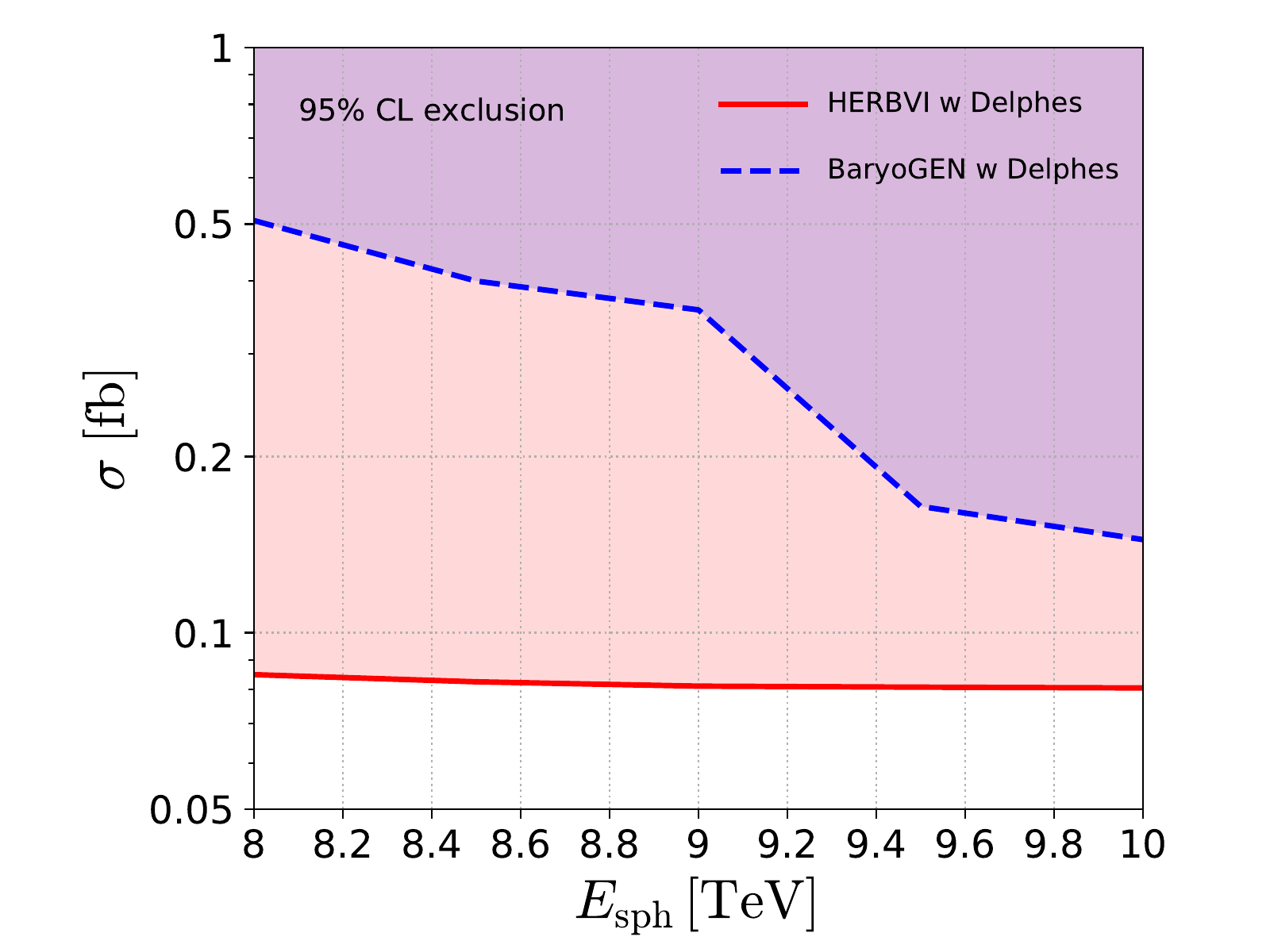}
    \hspace{-10mm}
    \includegraphics[width=0.52\textwidth,clip]{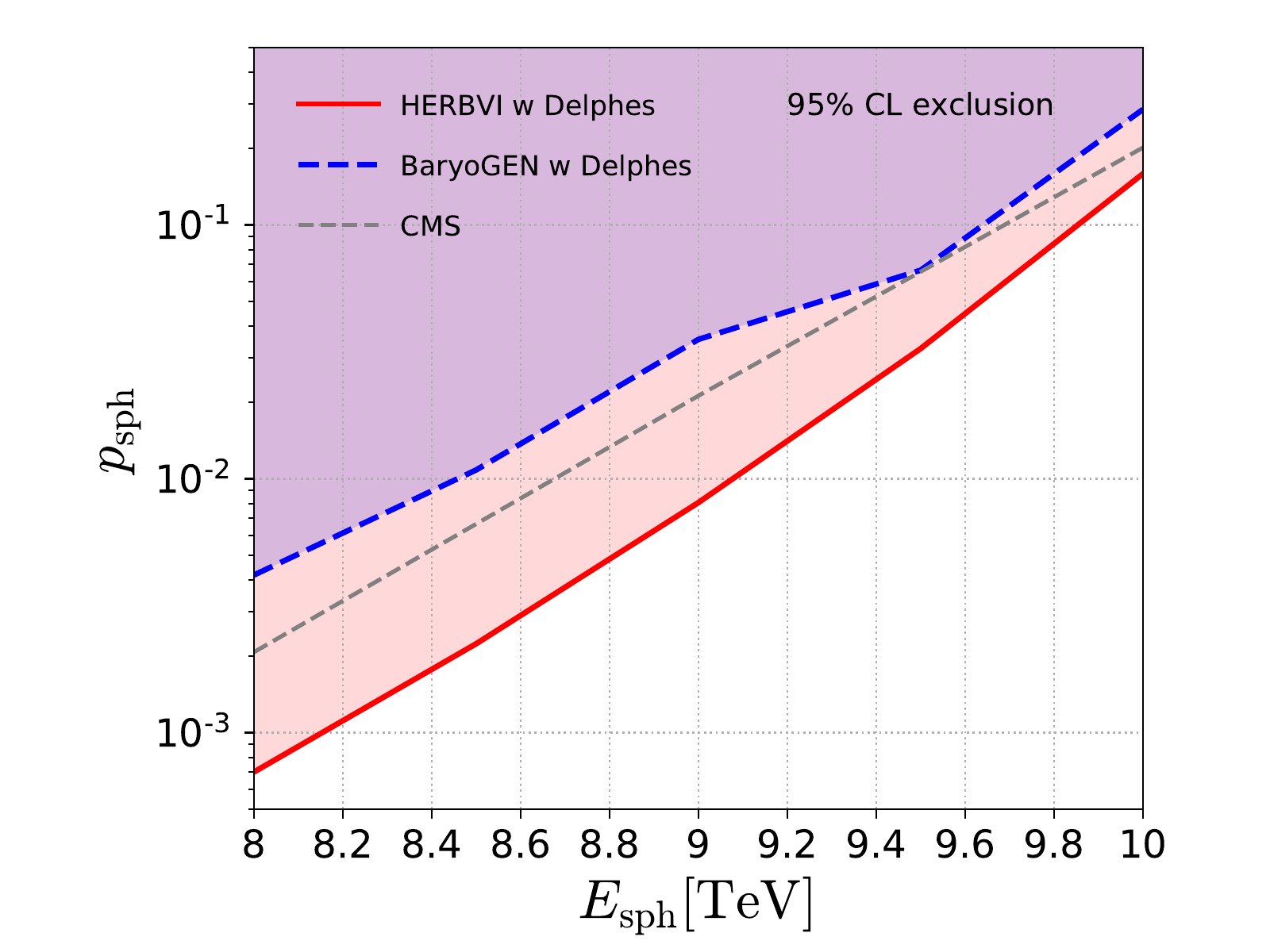}
    \caption{ 
    The 95\,\% CL exclusion limits obtained by recasting CMS black hole-sphaleron analysis \cite{cms}
    in the ($E_{\rm sph}$ vs $\sigma$) [left panel] and ($E_{\rm sph}$ vs $p_{\rm sph}$) [right panel] planes.
    The light-red and light-blue regions are excluded for {\tt HERBVI} and {\tt BaryoGEN}, respectively. 
    }
    \label{fig:limit}
\end{center}
\end{figure}

The left panel of Fig.~\ref{fig:limit} shows the $95 \%$ CL exclusion limits
in the ($E_{\rm sph}$ vs $\sigma$) plane.
The limit, $\sigma < \sigma_{\rm max}$, is obtained from
$\sigma_{\rm max} \cdot \epsilon^{(a^*)} \cdot L_{\rm int} = N_{\rm obs}^{{\rm max}(a^*)}$
with $L_{\rm int} = 35.9$ fb$^{-1}$.
One can see that
a stronger limit is placed on the multi-boson signal than on the zero-boson,
although CMS targets the latter in their analysis.
As discussed above, this is because the multi-boson processes give rise to larger multiplicity events,
enabling one to go to a large $N_{\rm min}$ region, 
where the Standard Model background is extremely small, without losing signal.
The cross-section limit is given by $\sigma_{\rm max} \simeq 0.08$ fb throughout the range of $E_{\rm sph}$ in the plot
for the multi-boson case.
This can be seen also in Table \ref{tb},
since the signal efficiency, $\epsilon^{(a^*)}$, and the observed upper limit, $N_{\rm obs}^{{\rm max}(a^*)}$, 
are almost independent of $E_{\rm sph}$ for this case.
Since $\epsilon^{(a^*)}$ and $N_{\rm obs}^{{\rm max}(a^*)}$ 
are close to their highest and lowest values, respectively,
there may be no room to further improve the limit for the multi-boson
case, except by increasing the integrated luminosity or the collider energy.

In the case of the zero-boson final state, 
the cross-section limit is stronger for larger $E_{\rm sph}$,
since the larger the $E_{\rm sph}$, the higher the $S_T$ on average. 
On the other hand, the hadronic cross-section decreases rapidly upon
increasing the partonic threshold energy, due to the PDF suppression 
for fixed $p_{\rm sph}$.
We show in the right panel of Fig.~\ref{fig:limit}
the $95 \%$ CL excluded regions
in the ($E_{\rm sph}$ vs $p_{\rm sph}$) plane.
As expected the limit on $p_{\rm sph}$ is weaker for larger $E_{\rm sph}$.
We superimpose in this plot the limit shown in the CMS analysis \cite{cms}.
This should be compared to the zero-boson limit 
obtained from our recasting of the analysis based on the fast detector simulation with {\tt Delphes}.
One can see that our limit is on the conservative side and 
reasonably close to the CMS limit.

\medskip
\section{Summary and conclusions}
\label{sec:conclusions}

In this paper we compared various kinematical and multiplicity distributions 
between the anomalous $B+L$ violating processes with zero- and multi-boson final states.
The former final state was assumed in the recent CMS analysis \cite{cms} 
on searching for instanton-induced processes at the LHC,
whilst many theoretical studies suggest that the latter is more realistic.  

We showed that there is a significant difference in jet multiplicity distributions;
the peak positions are around 10 for zero-boson and 22 for multi-boson final states,
for jets with $p_T > 30$ GeV.
We also found that a multi-boson event contains at least one lepton and photon 
with more than 50\,\% probability. 
On the other hand, the $S_T$ variable tends to be smaller for the multi-boson final state
because a large number of jets are rejected by the somewhat harsh $p_T$ cut ($p_T > 70$ GeV)
employed by the CMS analysis.

By applying the same event selection used in the CMS analysis \cite{cms}, 
we have derived a cross-section upper limit, $\sigma < 0.08$ fb, on the multi-boson processes
within the region we considered, $8 \le E_{\rm sph}/{\rm TeV} \le 10$. 
We found that unlike the zero-boson case, the most sensitive signal region is identified as 
the one with the highest jet multiplicity bins ($N_{\rm min} = 11$), in which the SM background 
is extremely low and less than 1 event for the current integrated luminosity, 35.9 fb$^{-1}$,
while the signal efficiency is very close to 100\%.
Therefore, this signal region almost maximizes the analysis efficiency in terms of $S/\sqrt{B}$
and further improvement cannot be expected within the current dataset. 
On the other hand, by employing even larger jet multiplicity bins, e.g.~$N_j \geq 15$,
one can expect some improvement on the analysis searching for the sphaleron-like process 
with multi-boson final states, for different energy or luminosity options.

\acknowledgments
We thank Greg Landsberg for helpful comments.
The work of KS is partially supported by the National Science Centre,
Poland, under research grants 2017/26/E/ST2/00135 and DEC-2016/23/G/ST2/04301.  
The work of BRW is partially supported by U.K. STFC consolidated grant ST/P000681/1.
KS and BRW are grateful for the hospitality of Kavli IPMU 
and the 4th Kavli IPMU-Durham IPPP-KEK-KIAS workshop ``Beyond the BSM'',
while part of this work was performed.
Kavli IPMU is supported by the World Premier International Research Center Initiative
(WPI Initiative), MEXT, Japan.

\addcontentsline{toc}{section}{References}
\bibliographystyle{JHEP}
\bibliography{ref}

\end{document}